\documentclass[letterpaper,english,pre, aps, reprint, longbibliography ]{revtex4-1}
\usepackage[T1]{fontenc}
\usepackage{babel}
\usepackage{units}
\usepackage{bm}
\usepackage{amsmath}
\usepackage{amssymb}
\usepackage{graphicx}
\usepackage[unicode=true,pdfusetitle,
 bookmarks=true,bookmarksnumbered=false,bookmarksopen=false,
 breaklinks=false,pdfborder={0 0 1},backref=false,colorlinks=false]
 {hyperref}



\begin{document}
\title{Exact scattering cross section for lattice-defect scattering of phonons
using the atomistic Green's function method}
\author{Zhun-Yong Ong}
\affiliation{Institute of High Performance Computing (IHPC), Agency for Science,
Technology and Research (A{*}STAR), 1 Fusionopolis Way, \#16-16 Connexis,
Singapore 138632, Republic of Singapore }
\email{ongzy@ihpc.a-star.edu.sg}

\date{\today}
\begin{abstract}
The use of structurally complex lattice defects, such as functional
groups, embedded nanoparticles, and nanopillars, to generate phonon
scattering is a popular approach in phonon engineering for thermoelectric
applications. However, the theoretical treatment of this scattering
phenomenon remains a formidable challenge, especially with regard
to the determination of the scattering cross sections and rates associated
with such lattice defects. Using the extended atomistic Green's function
(AGF) method, we describe how the numerically exact mode-resolved
scattering cross section $\sigma$ can be computed for a phonon scattered
by a single lattice defect. We illustrate the generality and utility
of the AGF-based treatment with two examples. In the first example,
we treat the isotopic scattering of phonons in a harmonic chain of
atoms . In the second example, we treat the more complex problem of
phonon scattering in a carbon nanotube (CNT) containing an encapsulated
C60 molecule which acts as a scatterer of the CNT phonons.  The application
of this method can enable a more precise characterization of lattice-defect
scattering and result in  a more controlled use of nanostructuring
and lattice defects in phonon engineering for thermoelectric applications.

\end{abstract}
\maketitle

\section{Introduction}

The deliberate use of defects to create inhomogeneities in the crystal
lattice is a known approach for reducing the lattice thermal conductivity
of a semiconducting material in thermoelectrics.~\citep{ZLiu:MRSB18_Nano}
The broken translational symmetry due to the defect results in heat-carrying
phonons undergoing phonon-defect scattering, which reduces the phonon
mean free path (MFP) and generates resistance to heat conduction.
This reduction in the MFP in turn depends on the type and size of
the defect and how phonons interact with it. In materials containing
a significant concentration of defects, the reduction in the thermal
conductivity can also be explained by phonon-defect scattering.~\citep{PKlemens:PPSA55_Scattering,JCallaway:PR59_Model,PKlemens:PR60_Thermal,MKlein:PR63_Phonon} 

Given the key role of phonon-defect scattering in heat conduction,~\citep{RGurunathan:PRAppl20_Analytical,HWei:CM23_Influence}
a theoretical description of the scattering process that takes into
account the atomistic structure of the defect is important. In such
a description, we associate the defect with a scattering cross section
$\sigma(\nu,\bm{q})$ that depends on the polarization $\nu$ and
wave vector $\bm{q}$ of the incident phonon. Physically, the scattering
cross section $\sigma$ corresponds to the size of the interaction
of the phonon with the defect and vanishes if the phonon does not
interact with the defect. It also determines the phonon-defect scattering
rates that are used as inputs in semiclassical phonon transport models.~\citep{JZiman:Book60_Electrons}
Therefore, the efficient computation of $\sigma$ is important for
understanding phonon-defect interactions and developing new approaches
in phonon engineering for thermoelectric applications. 

A popular atomistic approach to computing $\sigma$ is the $T$-matrix
method,~\citep{NMingo:PRB10_Cluster} which depends on the changes
in the interatomic force constants and atomic masses in the lattice
generated by the defect. In this method, the scattering cross section
is directly computed from the diagonal matrix element of the $T$
matrix.~\citep{EEconomou:Book83_Greens} Although this method has
been applied to local site defects such as isotopes, vacancies and
substitutional impurities,~\citep{AKundu:PRB11_Role,NProtik:PRB16_AbInitio,AKatre:PRL17_Exceptionally,CAPolanco:PRB18_AbInitio}
it remains challenging to extend it to a more general class of structurally
complex defects, such as functional groups,~\citep{JLee:JCP11_Single}
surface nanopillars,~\citep{HHonarvar:APL16_Thermal,RAnufriev:Nanoscale17_Aluminium,HHonarvar:PRB18_Two}
embedded nanoparticles,~\citep{JFeser:JAP19_Engineering,OChowdhury:JAP22_Phonon}
or molecular impurities,~\citep{NZuckerman:PRB08_Acoustic} because
of the additional degrees of freedom associated with such defects. 

In this paper, we propose an alternative approach for computing the
scattering cross section $\sigma$, based on the \emph{extended} atomistic
Green's function (AGF) method,~\citep{ZYOng:PRB15_Efficient,ZYOng:JAP18_Tutorial,ZYOng:PRB18_Atomistic}
which is commonly used to study ballistic phonon transport~\citep{NMingo:PRB03_Phonon,JSWang:PRB06_Nonequilibrium,WZhang:JHT07_Simulation,WZhang:NHT07_Atomistic,JSWang:EPJB08_Quantum,ZYOng:PRB15_Efficient}
and has also been applied to model elastic wave scattering~\citep{HKhodavirdi:IJMS24_Atomistic}
as well as to characterize the boundary roughness scattering of phonons.~\citep{ZYOng:PRB20_Structure,ZYOng:PRB24_Effect}
In the extended AGF method, which is a development of the original
AGF method introduced by Mingo and Yang for treating the total phonon
transmission in nanowires,~\citep{NMingo:PRB03_Phonon} mode-resolved
quantities, such as the scattering amplitudes as well as mode-resolved
transmission and reflection coefficients, can be calculated from the
scattering $S$ matrix.~\citep{ZYOng:PRB18_Atomistic} The method
for $\sigma$ is based on a rigorous interpretation of the scattering
amplitude and thus does not require any new computational technique
as $\sigma$ is derived from the diagonal element of the transmission
matrix computed within the extended AGF method.~\citep{ZYOng:PRB18_Atomistic}
For the sake of simplicity, we limit the discussion of our proposed
approach to quasi-one-dimensional systems, such as nanotubes or nanowires,
although the approach can be easily extended to two or three-dimensional
systems, which we briefly discuss in Appendix~\ref{sec:2D3DGeneralization}. 

Our proposed approach leverages on the existing AGF method and enables
the rapid computation of scattering cross sections and rates which
can be used to determine the reduction in phonon MFPs and facilitate
semiclassical phonon Boltzmann transport equation (BTE) modeling of
phonon transport in lattices with defects. Although defects destroy
the translational symmetry of the lattice, we can treat the defects
as randomly distributed, independent perturbations to the translationally
symmetric defect-free lattice, of which the eigenmodes are well-defined
as phonons. If we assume that the concentration of defects is dilute
enough for multiple scattering and interference effects between scatterers
to be neglected,~\citep{JZiman:Book60_Electrons,NMingo:PRB10_Cluster}
then the corresponding total defect scattering rate $\Gamma_{\text{defect}}$
for these phonons can be described by summing over the scattering
cross section and rate of each defect. In the phonon BTE, we can model
the reduction in the phonon lifetime and the lattice thermal conductivity,
because the inverse lifetime for the phonon is given by $\tau^{-1}=\Gamma_{\text{phonon}}+\Gamma_{\text{defect}}$,
where $\Gamma_{\text{phonon}}$ is the scattering rate due to anharmonic
phonon interactions. In systems with a higher concentration of defects,
we cannot ignore multiple scattering and alternative phonon transport
models should be used, because the effects of localization~\citep{PSheng:Book06_IntroductionWave}
may play a significant role in phonon transport.~\citep{ISavic:PRL08_CNTLocalization,AChaudhuri:PRB10_PhonLocalization}

Before we describe the methodology for computing $\sigma$, we give
an intuitive explanation of how it works. In the AGF framework, the
system is partitioned into the left lead, the device region containing
the scatterer, and the right lead. As an incoming phonon mode propagates
from one lead across the inhomogeneity within the device region and
towards the other lead, part of its flux is scattered by the inhomogeneity
while the remaining part of it is transmitted into the other lead.
As a result of the interaction with the inhomogeneity, the transmission
amplitude, which characterizes the outgoing phonon wave function,
acquires an additional phase factor in its amplitude, which we can
determine from the $S$ matrix. Because the $S$ matrix is unitary
and is essentially a bookkeeping device that connects the incoming
amplitudes to the outgoing amplitudes~\citep{ZYOng:PRB18_Atomistic},
this phase factor allows us to determine the \emph{change} in the
phonon wave function due to scattering, from which we can find the
scattering cross section $\sigma$ using the optical theorem. Since
we only examine the amplitude change in the outgoing phonon wave function
in the lead, we can thus treat the device as a \emph{black box} so
long as it has the same length as its homogeneous counterpart. This
treatment is more akin to the analysis of the asymptotic behavior
of the scattered wave function in scattering problems.~\citep{RNewton:Book82_Scattering,LBoya:PRA94_Optical}
It is more flexible as we are less constrained by the structure of
the complex defect, compared to the $T$ matrix approach where one
has to map the perturbed degrees of freedom to those of the homogeneous
system and to compute directly the $T$ matrix element associated
with the perturbation. 

The organization of this paper is as follows. We first describe the
theoretical basis for this method of computing $\sigma$. We derive
the key formula for the scattering cross section, given in Eq.~(\ref{eq:CrossSectionFormula}),
from the diagonal element of the transmission matrix. Next, we illustrate
the utility of the method using two examples. The first example is
the monoatomic harmonic chain with a single isotopic impurity while
the second example consists of the (10,10) armchair carbon nanotube
(CNT) with an encapsulated C60 molecule acting as the scatterer. In
the second example, we compute and analyze how the scattering cross
sections associated with the C60 molecule depend strongly on the polarization
and atomic motion of the incident phonon modes in the CNT, especially
for the low-frequency phonon subbands. 

\begin{figure}
\includegraphics[width=8cm]{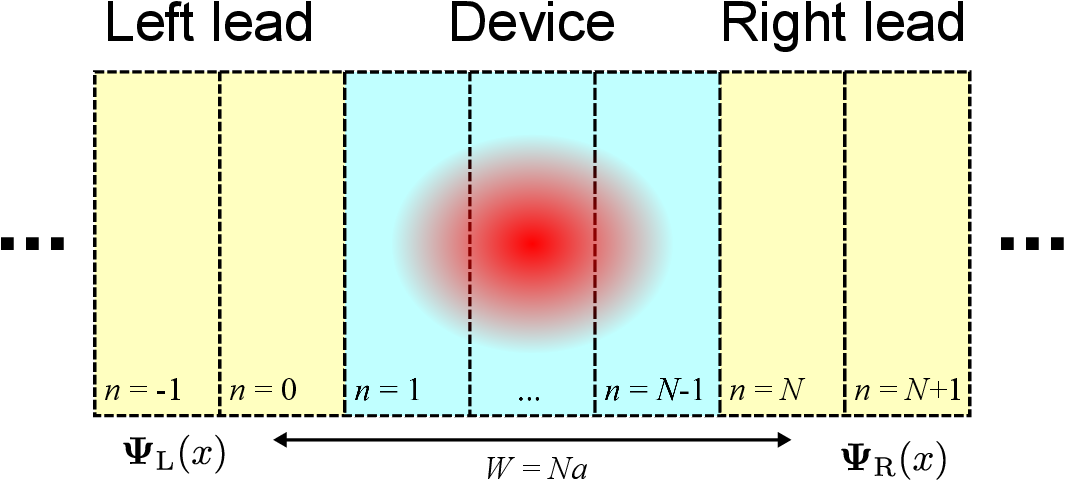}

\caption{Schematic of layered system in the AGF method. The inhomogeneity or
defect, which is represented by the red-shaded oval, is contained
within the layers $n=1$ to $N-1$ which correspond to the device
region. The distance between the left and right leads is $W=Na$. }

\label{fig:LayeredSystem}

\end{figure}

\section{Theoretical basis of AGF-based approach}

In this section, we derive the formula for the scattering cross section
by analyzing how the phonon wave function is perturbed by scattering
in one dimension. Related ideas are discussed in the pedagogical review
by Boya~\citep{LBoya:RNC08_Quantum} although they apply only to
continuous systems, such as the Schroedinger equation, and not to
the discrete lattice-type system that the AGF method is designed for. 

In the AGF method,~\citep{ZYOng:JAP18_Tutorial,ZYOng:PRB18_Atomistic}
which is outlined in Appendix~\ref{sec:ExtendedAGFMethod}, we arrange
the degrees of freedom in an infinite array of layers, which we enumerate
as $n=-\infty,\ldots,0,1,\ldots,\infty$ and assume are evenly spaced
with an interlayer distance of $a$, as shown in Fig.~\ref{fig:LayeredSystem}.
Thus, we can associate a discrete coordinate $x=na$ with the center
of the $n$-th layer. As before, we partition the system into three
subsystems: the left lead ($n\leq0$), the device ($1<n<N-1$), and
the right lead ($n\geq N$). In the left and right leads which are
homogeneous and identical to one another, each layer acts as a unit
cell, has $l$ degrees of freedom and is identical to the other cells
in the lead. In the device, the layers are not necessarily homogeneous
even though they are spaced evenly and can accommodate the additional
degrees of freedom associated with an inhomogeneity. In our derivation,
we discuss the scattering cross section associated with the inhomogeneity
for an  incoming phonon propagating rightward from the left lead towards
the device.

\subsection{Total outgoing phonon flux from scattering}

In the left lead ($n\leq0$), we associate the incident eigenwave
$\bm{\Psi}_{\text{inc}}(x)$ with a phonon mode of wave vector of
$k$, branch index $\nu$ ($1\leq\nu\leq l$), frequency $\omega_{\nu,k}$
and group velocity $v_{\nu,k}>0$. For convenience, we use $\nu k$
to label the phonon modes. We write the \emph{flux-normalized~}\footnote{By flux-normalized, we mean that the absolute square of the wave eigenfunction
gives us the flux, i.e., $j=|\Psi|^{2}=1$.} phonon `wave function' $\bm{\Psi}_{\text{L}}(x)$ in the left lead
as the sum of the incident and scattered waves:
\begin{equation}
\bm{\Psi}_{\text{L}}(x)=\bm{\Psi}_{\text{inc}}^{\text{L}}(x)+\bm{\Psi}_{\text{scatt}}^{\text{L}}(x)\ ,\label{eq:LeftLeadWave}
\end{equation}
where $\bm{\Psi}_{\text{inc}}^{\text{L}}(x)=\bm{\Phi}_{\nu k}(x)\equiv\sqrt{|v_{\nu k}|}\bm{u}_{\nu k}(x)e^{ikx}$
represents the incident $\nu k$ phonon mode~\footnote{A careful distinction should be made between the terms incident and
incoming. The former is used to refer to the unperturbed state when
the scatterer is absent while the latter is used to refer to the phonon
mode that is traveling towards the device region. A rightward-propagating
incident phonon consists of an incoming phonon eigenmode in the left
lead and an outgoing phonon eigenmode in the right lead.} with group velocity $v_{\nu k}$ and $\bm{u}_{\nu k}(x)$ denotes
its Bloch function satisfying the relationship $\bm{u}_{\nu k}(x)=\bm{u}_{\nu k}(x+a)$.
Because of flux normalization, Eq.~(\ref{eq:LeftLeadWave}) only
contains the contribution from the propagating modes and none from
the evanescent modes. In the absence of scattering, the phonon wave
function at any two layers at $x$ and $x^{\prime}$ only differ by
a phase factor determined by its wave vector, i.e., $\bm{\Psi}_{\text{L}}(x)=\bm{\Psi}_{\text{L}}(x^{\prime})e^{ik(x-x^{\prime})}$. 

The Bloch function $\bm{u}_{\nu k}(x)$ is a column vector whose $l$
elements correspond to the displacement degrees of freedom in a unit
cell. Thus, all the phonon wave functions are also column vectors.
On the righthand side (RHS) of Eq.~(\ref{eq:LeftLeadWave}), $\bm{\Psi}_{\text{scatt}}^{\text{L}}(x)$
represents the \emph{scattered} wave in the left lead and is given
by

\begin{equation}
\bm{\Psi}_{\text{scatt}}^{\text{L}}(x)=\sum_{\nu^{\prime}k^{\prime}}S(\nu^{\prime}k^{\prime},\nu k)\bm{\Phi}_{\nu^{\prime}k^{\prime}}(x)\Theta(-v_{\nu^{\prime}k^{\prime}})\ ,\label{eq:LeftLeadScatteredWave}
\end{equation}
which is written as a sum of the \emph{reflected} leftward-propagating
flux-normalized $\nu^{\prime}k^{\prime}$ eigenmodes associated with
the scattering amplitudes $S(\nu^{\prime}k^{\prime},\nu k)$.\footnote{The Heaviside term $\Theta(-v_{k^{\prime}})$ eliminates the contributions
from the transmitted rightward-propagating waves in Eq.~(\ref{eq:LeftLeadScatteredWave}).} The sum over $\nu^{\prime}k^{\prime}$ in Eq.~(\ref{eq:LeftLeadScatteredWave})
is over all possible branches and wave vectors at the incident phonon
frequency $\omega_{\nu k}$, i.e., $\omega_{\nu^{\prime}k^{\prime}}=\omega_{\nu k}$.
We can associate two fluxes with the left-lead wave function: one
from the incident mode and the other from the modes reflected from
the device into the left lead. The normalized incident flux is $j_{\text{inc}}^{\text{L}}(\nu k)=1$
while the reflected flux is 
\begin{equation}
j_{\text{scatt}}^{\text{L}}(\nu k)=\sum_{\nu^{\prime}k^{\prime}}|S(\nu^{\prime}k^{\prime},\nu k)|^{2}\Theta(-v_{\nu^{\prime}k^{\prime}})\ .\label{eq:LeftLeadScatteredFlux}
\end{equation}

In the right lead ($n\geq N$), we can likewise write the \emph{flux-normalized}
phonon wave function as the sum of transmitted rightward-propagating
fluxes, i.e.,

\begin{equation}
\bm{\Psi}_{\text{R}}(x)=\sum_{\nu^{\prime}k^{\prime}}S(\nu^{\prime}k^{\prime},\nu k)\bm{\Phi}_{\nu^{\prime}k^{\prime}}(x)e^{-ik^{\prime}W}\Theta(v_{\nu^{\prime}k^{\prime}})\ .\label{eq:RightLeadScatteredWave}
\end{equation}
The RHS of Eq.~(\ref{eq:RightLeadScatteredWave}) contains the incident
and outgoing forward-scattered wave functions. Each summand has a
phase factor of $e^{-ik^{\prime}W}$, where $W$ denotes the width
of the scatterer and is taken as the distance between the lead edges
(from $n=0$ in the left lead to $n=N$ in the right lead), so we
have $W=Na$. In the absence of a scatterer (e.g. homogeneous waveguide),
the forward-scattering amplitude in the right lead is $\overline{S}(\nu^{\prime}k^{\prime},\nu k)=\delta_{\nu^{\prime}\nu}\delta_{k^{\prime},k}e^{ik^{\prime}W}$
where $\delta_{k^{\prime},k}$ denotes the Kronecker delta function,
and we recover the solution $\overline{\bm{\Psi}}_{\text{R}}(x)=\sum_{\nu^{\prime}k^{\prime}}\overline{S}(\nu^{\prime}k^{\prime},\nu k)\bm{\Phi}_{\nu^{\prime}k^{\prime}}e^{-ik^{\prime}W}\Theta(v_{\nu^{\prime}k^{\prime}})=\bm{\Phi}_{\nu k}(x)$,
which is just the incident eigenmode. 

Like in Eq.~(\ref{eq:LeftLeadWave}), we can express Eq.~(\ref{eq:RightLeadScatteredWave})
as the sum of the incident and scattered wave function or 
\begin{equation}
\bm{\Psi}_{\text{R}}(x)=\bm{\Psi}_{\text{inc}}^{\text{R}}(x)+\bm{\Psi}_{\text{scatt}}^{\text{R}}(x)\ ,\label{eq:RightLeadWave}
\end{equation}
where $\bm{\Psi}_{\text{inc}}^{\text{R}}(x)=\overline{\bm{\Psi}}_{\text{R}}(x)$
is the incident eigenwave. This implies that the scattered wave function
in the right lead is the remainder or $\bm{\Psi}_{\text{scatt}}^{\text{R}}(x)=\bm{\Psi}_{\text{R}}(x)-\bm{\Psi}_{\text{inc}}^{\text{R}}(x)=\sum_{\nu^{\prime}k^{\prime}}[S(\nu^{\prime}k^{\prime},\nu k)-\overline{S}(\nu^{\prime}k^{\prime},\nu k)]\bm{\Phi}_{\nu^{\prime}k^{\prime}}(x)e^{-ik^{\prime}W}\Theta(v_{\nu^{\prime}k^{\prime}})$.
Therefore, the associated flux for $\bm{\Psi}_{\text{scatt}}^{\text{R}}(x)$,
which represents the scattered flux exiting into the right lead, is
\begin{equation}
j_{\text{scatt}}^{\text{R}}(\nu k)=\sum_{\nu^{\prime}k^{\prime}}|S(\nu^{\prime}k^{\prime},\nu k)-\overline{S}(\nu^{\prime}k^{\prime},\nu k)|^{2}\Theta(v_{\nu^{\prime}k^{\prime}})\ .\label{eq:RightLeadScatteredFlux}
\end{equation}
The total outgoing flux from scattering in the left and right leads
is the sum of the outgoing scattered fluxes, i.e., $j_{\text{scatt}}(\nu k)=j_{\text{scatt}}^{\text{L}}+j_{\text{scatt}}^{\text{R}}=\sum_{\nu^{\prime}k^{\prime}}|S(\nu^{\prime}k^{\prime},\nu k)-\overline{S}(\nu^{\prime}k^{\prime},\nu k)|^{2}$,
which can be expanded to yield a sum with three terms in the summand,
i.e.,

\begin{align*}
j_{\text{scatt}}(\nu k)= & \sum_{\nu^{\prime}k^{\prime}}\{|S(\nu^{\prime}k^{\prime},\nu k)|^{2}+|\overline{S}(\nu^{\prime}k^{\prime},\nu k)|^{2}\\
 & -2\text{Re}[S(\nu^{\prime}k^{\prime},\nu k)\overline{S}(\nu^{\prime}k^{\prime},\nu k)^{*}]\}\ .
\end{align*}
The sum over each of the first two terms yields a value of unity because
the unitarity of the $S$ matrix implies that $\sum_{\nu^{\prime}k^{\prime}}|S(\nu^{\prime}k^{\prime},\nu k)|^{2}=\sum_{\nu^{\prime}k^{\prime}}|\overline{S}(\nu^{\prime}k^{\prime},\nu k)|^{2}=1$.
The sum over the remaining term yields $-2\text{Re}[S(\nu k,\nu k)\overline{S}(\nu k,\nu k)^{*}]$,
given that $\overline{S}(\nu^{\prime}k^{\prime},\nu k)=0$ for $\nu^{\prime}k^{\prime}\neq\nu k$.
Hence, we obtain

\begin{equation}
j_{\text{scatt}}(\nu k)=2\text{Re}\left[1-\frac{S(\nu k,\nu k)}{\overline{S}(\nu k,\nu k)}\right]\ .\label{eq:FinalTotalScatteringFlux}
\end{equation}

The expression in Eq.~(\ref{eq:FinalTotalScatteringFlux}) relates
the total outgoing flux from scattering to the transmission amplitude
$S(\nu k,\nu k)$. In the absence of any scatterer, Eq.~(\ref{eq:FinalTotalScatteringFlux})
vanishes as expected since $S(\nu k,\nu k)=\overline{S}(\nu k,\nu k)$.
Although this is not immediately obvious, Eq.~(\ref{eq:FinalTotalScatteringFlux})
is just a restatement of the well-known \emph{optical theorem} for
a one-dimensional multimodal system. For convenience, we define the
\emph{forward-scattering amplitude} $\mathcal{F}(\nu k)$ via the
relationship $S(\nu k,\nu k)=\overline{S}(\nu k,\nu k)[1+\mathcal{F}(\nu k)]$
or 
\begin{equation}
\mathcal{F}(\nu k)=S(\nu k,\nu k)e^{-ikW}-1\ .\label{eq:ForwardScatteringAmplitude}
\end{equation}
This allows us to simplify Eq.~(\ref{eq:FinalTotalScatteringFlux})
to obtain the compact expression $j_{\text{scatt}}(\nu k)=-2\text{Re}[\mathcal{F}(\nu k)]$.

\subsection{Scattering cross section from the transmission-matrix diagonal element}

Given that the incident wave function is $\bm{\Psi}_{\text{inc}}(x)=\bm{\Phi}_{\nu k}(x)$,
the incoming flux for the incident $\nu k$ mode is $j_{\text{inc}}=1$.
Therefore, the scattering cross section associated with the scatterer
between the two leads is 
\begin{equation}
\sigma(\nu k)=\frac{j_{\text{scatt}}}{j_{\text{inc}}}A=-2A\text{Re}[\mathcal{F}(\nu k)]\ ,\label{eq:TotalCrossSection}
\end{equation}
where $A$ denotes the\emph{ geometrical cross section} of the system.
It is worth remarking that the expression in Eq.~(\ref{eq:TotalCrossSection})
agrees with that obtained by Boya and Murray~\citep{LBoya:PRA94_Optical}
for quantum-mechanical scattering in a continuous system, even though
the derivation leading to Eq.~(\ref{eq:TotalCrossSection}) is purely
classical. This suggests that the generalized optical theorem obtained
in Ref.~\citep{LBoya:PRA94_Optical} can be extended to non-quantum-mechanical
settings.

In the absence of a scatterer, Eq.~(\ref{eq:TotalCrossSection})
yields $\sigma=0$ as expected since $S(\nu k,\nu k)=\overline{S}(\nu k,\nu k)$
and $\mathcal{F}(\nu k)=0$. On the other hand, we note that $\sigma\leq4A$,
i.e., the maximum value of the total scattering cross section can
be as large as \emph{four times} the geometrical cross section when
$S(\nu k,\nu k)=-\overline{S}(\nu k,\nu k)$ and $\mathcal{F}(\nu k)=-2$.~\citep{LBoya:RNC08_Quantum}
While it appears to be counterintuitive and even possibly unphysical,
this result also occurs in quantum mechanics for scattering by a hard
sphere in three dimensions.~\citep{JSakurai:Book20_ModernQuantumMechanics}
This is a consequence of our partitioning of the wave function in
$\bm{\Psi}_{\text{L}}(x)$ and $\bm{\Psi}_{\text{R}}(x)$ into two
\emph{non-orthogonal} components: the incident wave $\bm{\Psi}_{\text{inc}}$
and the scattered wave $\bm{\Psi}_{\text{scatt}}$ which represents
the part of the original wave that has been perturbed. 
\begin{widetext}
To find $S(\nu k,\nu k)$ for the $\nu k$ mode in the left lead,
we use the AGF method to compute the $l\times l$ transmission matrix
for the inhomogeneous waveguide, given by~\citep{ZYOng:JAP18_Tutorial,ZYOng:PRB18_Atomistic}
\begin{equation}
\boldsymbol{t}_{\text{RL}}(\omega)=\frac{2i\omega}{a}[\boldsymbol{V}_{\text{R}}^{\text{ret}}(+)]^{1/2}[\boldsymbol{U}_{\text{R}}^{\text{ret}}(+)]^{-1}\boldsymbol{G}_{\text{RL}}^{\text{ret}}[\boldsymbol{U}_{\text{L}}^{\text{adv}}(-)^{\dagger}]^{-1}[\boldsymbol{V}_{\text{L}}^{\text{adv}}(-)]^{1/2}\ ,\label{eq:TransmissionMatrix}
\end{equation}
where $\omega=\omega_{\nu k}$ and $\boldsymbol{G}_{\text{RL}}^{\text{ret}}$
denotes the Green's function between the rightmost left-lead ($n=0$)
and the leftmost right-lead ($n=N$) layers. In Eq.~(\ref{eq:TransmissionMatrix}),$\boldsymbol{V}_{\text{R}}^{\text{ret}}(+)$
and $\boldsymbol{V}_{\text{L}}^{\text{adv}}(-)$ correspond to the
velocity matrices for the outgoing right-lead phonons and the incoming
left-lead phonons, respectively, while $\boldsymbol{U}_{\text{R}}^{\text{ret}}(+)$
and $\boldsymbol{U}_{\text{L}}^{\text{adv}}(-)$ correspond to the
outgoing right-lead eigenmodes and the incoming left-lead eigenmodes,
respectively. A more detailed description of these matrices is given
in Eqs.~(\ref{eq:BlochMatrixEigenmodes})-(\ref{eq:LeftGoingVelocityMatrix})
in Appendix~\ref{sec:ExtendedAGFMethod}. The transmission matrix
element $[\boldsymbol{t}_{\text{RL}}(\omega)]_{pq}$ connects the
flux amplitude of the $p$-th outgoing right-lead modes to the $q$-th
incoming left-lead modes for $p,q=1,\ldots,l$. Because the left and
right leads are identical, the set of incoming \emph{extended} left-lead
modes, as indexed by $\nu k$, is the same as the set of outgoing
extended right-lead modes, and we can order the extended phonon modes
by their wave vectors such that $k_{1}<k_{2}<\ldots$. 
\end{widetext}

If we discard the evanescent modes and order the row and column indices
according to the size of $k$, then we can express Eq.~(\ref{eq:TransmissionMatrix})
as a $\bar{l}\times\bar{l}$ matrix
\begin{equation}
\overline{\boldsymbol{t}}_{\text{RL}}(\omega)=\left(\begin{array}{ccc}
S(\nu_{1}k_{1},\nu_{1}k_{1}) & \cdots & S(\nu_{1}k_{1},\nu_{\bar{l}}k_{\bar{l}})\\
\vdots & \ddots & \vdots\\
S(\nu_{\bar{l}}k_{\bar{l}},\nu_{1}k_{1}) & \cdots & S(\nu_{\bar{l}}k_{\bar{l}},\nu_{\bar{l}}k_{\bar{l}})
\end{array}\right)\ ,\label{eq:RationalizedTransmissionMatrix}
\end{equation}
where $\bar{l}\leq l$ is the number of\emph{ extended propagating}
incoming or outgoing phonon modes at frequency $\omega$. Using Eqs.~(\ref{eq:ForwardScatteringAmplitude})
and (\ref{eq:TotalCrossSection}), we obtain the scattering cross
section for the incoming $\nu_{p}k_{p}$ phonon ($1\leq p\leq\bar{l}$):
\begin{equation}
\sigma(\nu_{p}k_{p})=2A\text{Re}(1-[\overline{\boldsymbol{t}}_{\text{RL}}(\omega)]_{pp}e^{-ik_{p}W})\ .\label{eq:CrossSectionFormula}
\end{equation}
In the absence of any scattering, the expression for $\sigma(\nu_{p}k_{p})$
in Eq.~(\ref{eq:CrossSectionFormula}) vanishes as expected, because
$[\overline{\boldsymbol{t}}_{\text{RL}}(\omega)]_{pp}=e^{ik_{p}W}$
which is equal to the phase change of the phonon wave function as
it moves from layer $n=0$ to layer $n=N$. 

Here, we note one limitation of our method for calculating the scattering
cross section. In Eq.~(\ref{eq:TransmissionMatrix}), the transmission
matrix depends on the phonon group velocities. Hence, the scattering
cross section is ill-defined for eigenmodes that have zero group velocity
and cannot propagate spatially.

\section{Examples}

\subsection{Monoatomic harmonic chain with a single isotopic impurity}

\begin{figure}
\includegraphics[width=8cm]{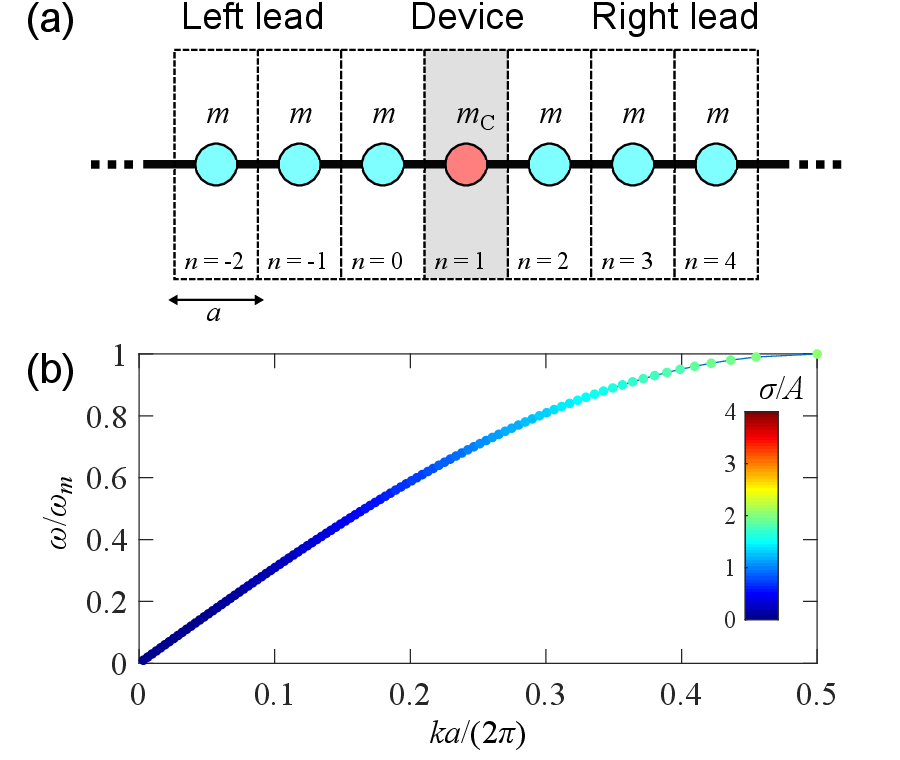}

\caption{(a) Schematic of monoatomic harmonic chain with a single isotopic
impurity $m_{c}$. (b) Plot of the $\omega-k$ phonon dispersion for
the single phonon branch and the normalized scattering cross section
$\sigma/A$ values of the phonon eigenmodes, with values indicated
by the color bar, for $\epsilon=1$. }

\label{fig:MonoatomicHarmonicChain}

\end{figure}

To illustrate the AGF method for calculating the scattering cross
section, we apply it to the simple archetypal example~\citep{PAllen:PRB22_Heat}
of phonon scattering by a single isotopic impurity of mass $m_{\text{C}}$
in a monoatomic harmonic chain with an interatomic distance of $a$,
as shown in Fig\@.~\ref{fig:MonoatomicHarmonicChain}(a). In this
system, we treat the isotopic impurity as the device; the mass and
interatomic force constants of the rest of the system are given by
$m$ and $\zeta$, respectively. For the monoatomic harmonic chain,
the phonon dispersion is given by $\omega(k)^{2}=\omega_{m}^{2}\sin^{2}(ka/2)$,
where $\omega_{m}=2\sqrt{\zeta/m}$, and the phonon group velocity
is given by $v(\omega)=\pm\frac{1}{2}a(\omega_{m}^{2}-\omega^{2})^{1/2}$.
As there is only one phonon branch, all the terms in Eq.~(\ref{eq:TransmissionMatrix})
are scalars. Thus, Eq.~(\ref{eq:TransmissionMatrix}) simplifies
to $t_{\text{RL}}(\omega)=i\omega(\omega_{m}^{2}-\omega^{2})^{1/2}G_{\text{RL}}^{\text{ret}}(\omega)$.

\subsubsection{Analytical expression for Green's function}

At frequency $\omega<\omega_{m}$ , the Green's function of the device
is given by~\citep{WZhang:NHT07_Atomistic} 
\[
G_{\text{C}}^{\text{ret}}(\omega)=[\omega^{2}+i0^{+}-\frac{2\zeta}{m_{\text{C}}}-\Sigma_{\text{L}}^{\text{ret}}(\omega)-\Sigma_{\text{R}}^{\text{ret}}(\omega)]^{-1}\ ,
\]
where $\Sigma_{\text{L}/\text{R}}^{\text{ret}}(\omega)=\frac{\zeta^{2}}{mm_{\text{C}}}g(\omega)$
denotes the self-energy term associated with the left/right lead and
$g(\omega)=\frac{1}{2}(\omega^{2}+i0^{+}-\frac{2\zeta}{m})-\frac{1}{2}[(\omega^{2}+i0^{+}-\frac{2\zeta}{m})^{2}-4(\frac{\zeta}{m})^{2}]^{1/2}$.
The analytical expression for the Green's function $G_{\text{RL}}^{\text{ret}}(\omega)$,
which describes the correlation between the ends of the left and right
leads, is given by $G_{\text{RL}}^{\text{ret}}(\omega)=g(\omega)H_{\text{RC}}G_{\text{C}}^{\text{ret}}(\omega)H_{\text{CL}}g(\omega)$,~\citep{ZYOng:PRB15_Efficient,ZYOng:JAP18_Tutorial,ZYOng:PRB18_Atomistic}
where $H_{\text{RC}}=H_{\text{CL}}=\zeta/\sqrt{mm_{\text{C}}}$ denotes
the lead-device coupling term. 

\subsubsection{Scattering cross section and rate}

After some algebra, we obtain from Eq.~(\ref{eq:CrossSectionFormula})
the expression for the scattering cross section of the single isotopic
impurity, normalized by the geometrical cross section $A$,
\begin{equation}
\frac{\sigma(\omega)}{A}=\frac{2\epsilon^{2}\omega^{4}}{\epsilon^{2}\omega^{4}+\omega^{2}(\omega_{m}^{2}-\omega^{2})}\ ,\label{eq:SingleImpurityCrossSection}
\end{equation}
where $\epsilon=(m_{\text{C}}-m)^{2}/m^{2}$ is the dimensionless
parameter for the mass difference. In Eq.~(\ref{eq:SingleImpurityCrossSection}),
$\sigma/A$, which increases monotonically with $\omega$ and is shown
in Fig.~\ref{fig:MonoatomicHarmonicChain}(b) for $\epsilon=1$,
has a value between 0 (no scattering) and 2 (total scattering) and
never reaches the theoretical upper bound of 4 from Eq.~(\ref{eq:TotalCrossSection}).
The scattering rate is 
\begin{equation}
\Gamma(\omega)=\frac{N_{i}\sigma_{\text{tot}}(\omega)v(\omega)}{NaA}=n_{i}\frac{\epsilon\omega^{4}(\omega_{m}^{2}-\omega^{2})^{1/2}}{\epsilon\omega^{4}+\omega^{2}(\omega_{m}^{2}-\omega^{2})}\ ,\label{eq:HarmonicChainScatteringRate}
\end{equation}
where $n_{i}=\frac{N_{i}}{Na}$ denotes the lineal concentration of
isotopic impurities. In the $\epsilon\ll1$ limit, Eq.~(\ref{eq:HarmonicChainScatteringRate})
reduces to 
\[
\Gamma(\omega)=\frac{n_{i}\epsilon\omega^{2}}{(\omega_{m}^{2}-\omega^{2})^{1/2}}+O(\epsilon^{2})\ ,
\]
in agreement with the result~\footnote{We specifically refer to the expression given by Eq.~(A.3) in Ref.~\citep{PAllen:PRB22_Heat}. }
obtained in Ref.~\citep{PAllen:PRB22_Heat}.

\subsection{Carbon nanotube with an encapsulated C60 molecule}

\begin{figure}
\includegraphics[width=8cm]{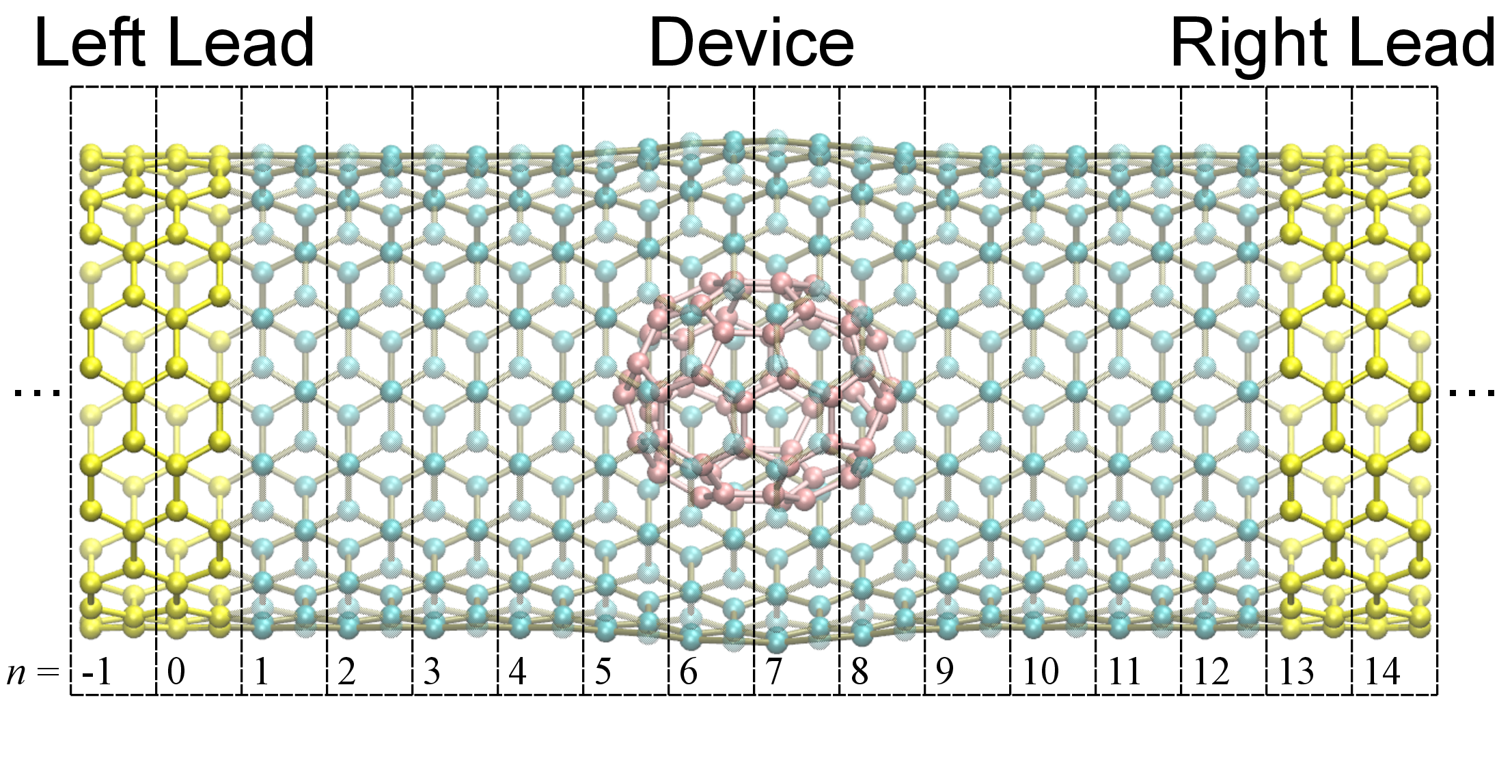}

\caption{Schematic of the layer arrangement in the (10,10) armchair carbon
nanotube with a C60 molecule (pink-shaded atoms) encapsulated within
the device region (turquoise-shaded atoms) in layers $n=1$ to $12$.
The layers in the leads, which extend semi-infinitely to the left
or right, are represented by the yellow-shaded atoms. }

\label{fig:C60CNT}
\end{figure}

Next, we apply the AGF method to the more complex problem of a single
C60 molecule encapsulated in the center of a (10,10) armchair CNT,
as shown in Fig.~\ref{fig:C60CNT}. The purpose of our discussion,
which is not meant to be comprehensive, is to illustrate how the scattering
cross section depends on the properties of the phonon mode and its
interaction with the \emph{external} scatterer. Unlike the monoatomic
chain in which the atoms can only move along the longitudinal axis
and there is only one phonon branch, the CNT has 120 phonon branches
because of the 40 carbon atoms in the unit cell and the three degrees
of freedom in atomic motion. These differences in the atomic motion
of the phonons affect their interaction with the C60 molecule. In
the system shown in Fig.~\ref{fig:C60CNT}, the finite CNT region
encapsulating the C60 molecule acts as the device while each lead
comprises a semi-infinite homogeneous CNT, in which each layer corresponds
to the unit cell. Because the C60-CNT interaction breaks the translational
symmetry of the interatomic forces in the CNT, the incoming phonons
from the leads are scattered and we can associate a scattering cross
section for each phonon mode of the CNT. 

\begin{figure}
\includegraphics[width=8cm]{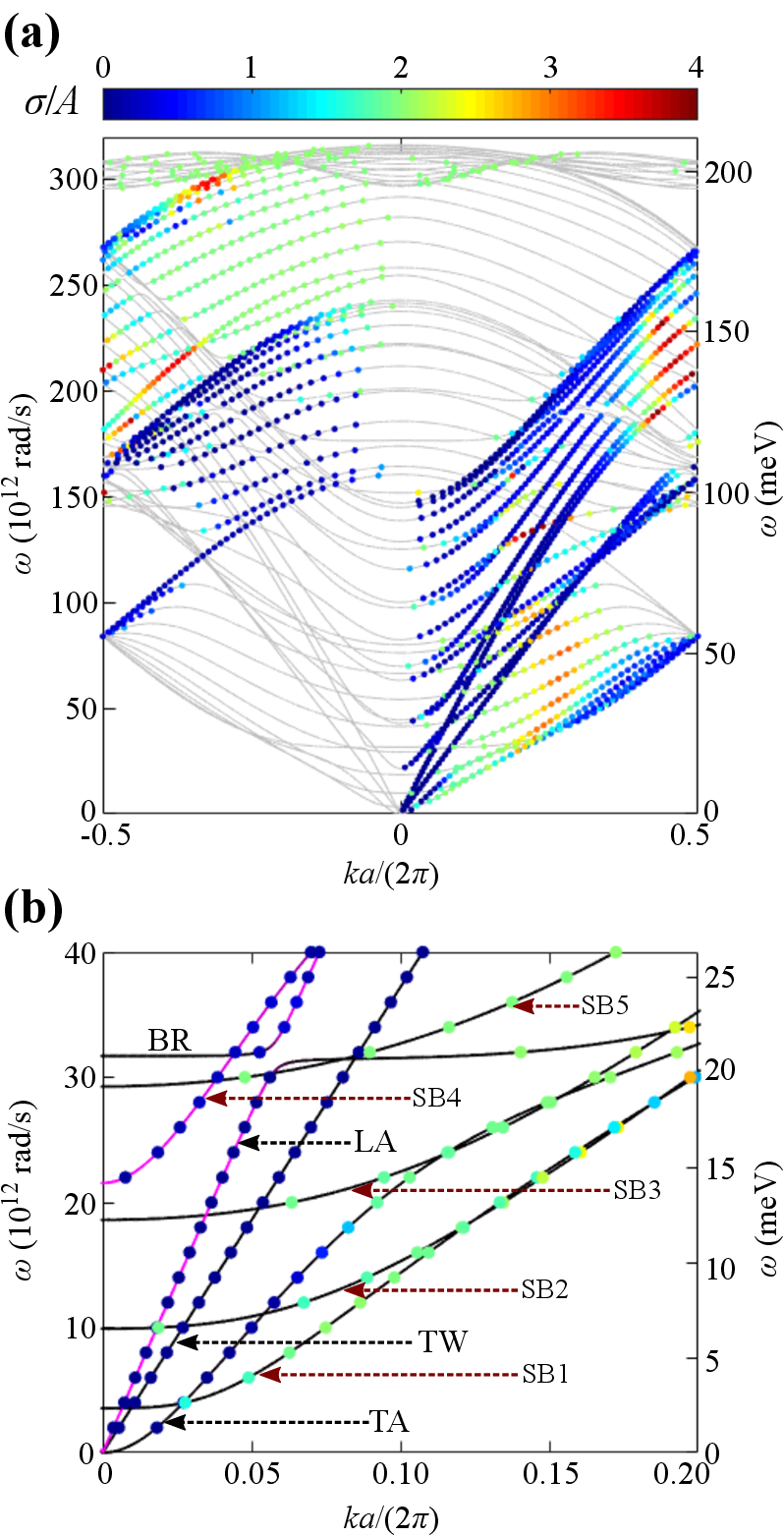}

\caption{(a) Plot of scattering cross section $\sigma/A$ for each rightward-propagating
phonon mode (solid circle) in the (10,10) armchair CNT, distributed
over the 66 distinct phonon branches. The values of $\sigma/A$ are
between 0 and 4 and are indicated by color according to the color
bar. (b) The corresponding plot at low frequencies ($\omega<4\times10^{13}$
rad/s). We observe the three distinct acoustic phonon branches (LA,
TW and TA) and the radial breathing mode (BR) as well as the five
lowest phonon subbands which are indicated by brown dashed arrows
and labeled SB1 to SB5. The phonon branches are also colored according
to the percentage of their eigenmode atomic displacements parallel
to the longitudinal axis, with the color varying continuously between
bright magenta (100\%) and black (0\%).}

\label{fig:DispersionCrossSection}
\end{figure}

We model the covalent C-C bonds within the CNT or C60 molecule using
the Tersoff potential optimized for graphene~\citep{LLindsay:PRB10_Optimized}
and the weak van der Waals forces between the CNT and the C60 molecule
using the Lennard-Jones (LJ) potential with parameters from Ref.~\citep{MOhnishi:PRB21_Strain}.
In Fig.~\ref{fig:C60CNT}, the device region spans 12 layers ($n=1$
to $12$) in the CNT as the C60 molecule couples to the atoms in them
through the LJ potential. We also observe a small bulge in the CNT
in the device region from the C60-CNT interaction. We optimize the
structure using GULP~\citep{JGale:MolSim03_gulp} and extract the
interatomic force constants needed as inputs for the AGF computations,
which are described in Refs.~\citep{ZYOng:JAP18_Tutorial,ZYOng:PRB18_Atomistic},
and the phonon dispersion of the CNT shown in Fig.~\ref{fig:DispersionCrossSection}(a). 

\subsubsection{Distribution of mode-dependent scattering cross sections}

We compute the area-normalized scattering cross section $\sigma(\nu k)/A$
for each rightward-propagating $\nu k$ phonon mode in the (10,10)
armchair CNT using Eqs.~(\ref{eq:TransmissionMatrix}) and (\ref{eq:CrossSectionFormula})
for a range of angular frequencies $\omega$ in intervals of $\Delta\omega=2\times10^{12}$
rad/s from $\omega=2\times10^{12}$ to $3.16\times10^{14}$ rad/s.
At each $\omega$, we obtain a set of phonon modes with the wave vectors
$k_{1},k_{2},\ldots$, which we plot as solid circles in Fig.~\ref{fig:DispersionCrossSection}.
As in Fig.~\ref{fig:MonoatomicHarmonicChain}, we indicate the value
of $\sigma/A$ of each mode by the color of its solid circle symbol.
For ease of interpretation, the solid circle symbols are superimposed
on the $\omega-k$ phonon dispersion curves for the (10,10) armchair
CNT.~\citep{MDresselhaus:AdvPhys00_Phonons}

In Fig.~\ref{fig:DispersionCrossSection}(a), there are 66 distinct
ones due to the degeneracies in the (10,10) armchair CNT.~\citep{MDresselhaus:AdvPhys00_Phonons}
At $k=0$ in the $\omega\rightarrow0$ limit, there are four acoustic
phonon branches corresponding to the longitudinal acoustic (LA), the
azimuthal twisting (TW) and the doubly degenerate radial transverse
acoustic (TA) phonons.~\citep{GMahan:PRB02_Oscillations,LChico:PRB06_LowFrequency}
In addition, there is a breathing mode (BR) associated with the radial
oscillation of the CNT at $k=0$. The remaining phonon branches consist
of subbands with $\omega>0$ at $k=0$. We observe two distinct $\sigma/A$
trends for the lower-frequency phonon subbands in Fig.~\ref{fig:DispersionCrossSection}(a).
In the subbands with higher group velocities and near-linear dispersion,
$\sigma/A$ is vanishingly small because the atomic motion for the
phonon modes is primarily along the longitudinal direction~\citep{MDresselhaus:AdvPhys00_Phonons}.
On the other hand, in the subbands with smaller group velocities and
a more parabolic dispersion, $\sigma/A$ increases significantly with
$\omega$, indicating stronger interaction with the C60 molecule.
For the subbands correspond to the high-frequency optical phonons,
we have $\sigma/A\sim2$, which indicates strong scattering. 

\subsubsection{Low-frequency phonon subbands}

For a more detailed analysis of how the C60 molecule scatters different
phonon modes, the low-frequency $\sigma/A$ distribution is shown
in Fig.~\ref{fig:DispersionCrossSection}(b). The individual phonon
dispersion curves for the acoustic phonons (LA, TW and TA) and the
five lowest phonon subbands (SB1 to SB5) are also displayed, with
the color of the individual curves indicating the percentage of the
eigenmode atomic displacements parallel to the longitudinal axis.
We observe that the $\sigma/A$ values for the acoustic phonons approaches
0 in the long wavelength ($k=0$) limit and increases monotonically
with $k$, similar to the results for the monoatomic chain in Fig.~\ref{fig:MonoatomicHarmonicChain}.
In particular, the LA and TW phonons, whose atomic motions are in
the longitudinal and azimuthal directions, respectively, have very
small $\sigma/A$ values which we can attribute to their weak coupling
to the C60 molecule. The TA phonons however see a significant increase
in $\sigma/A$ as $k$ increases because of the radial motion of their
modes.

On the other hand, the modes in the low-frequency phonon subbands
(SB1, SB2, SB3 and SB5) have a finite $\sigma/A$ in the $k=0$ limit
and increases with $\omega$. This implies that that these modes,
which are polarized in the plane normal to the longitudinal axis and
whose atomic motions are mainly in the radial and azimuthal directions,~\citep{GMahan:PRB02_Oscillations,LChico:PRB06_LowFrequency,MOhnishi:PRB21_Strain}
are sensitive to interaction with the C60 molecule. However, in the
SB4 subband, the low $\sigma/A$ values indicate that its phonon modes,
which are longitudinally polarized like those in the LA branch, interact
weakly with the C60 molecule. This observation is consistent with
the results for carbon nanotube peapods in Ref.~\citep{MOhnishi:PRB21_Strain}
where the authors report a weak reduction in the low-frequency LA
phonon lifetimes. The results in Fig.~\ref{fig:DispersionCrossSection}(b)
support the finding that the phonon scattering by the C60 molecule
is strongly dependent on phonon polarization, with phonon modes whose
atomic motions have a radial component being more strongly scattered
by the C60 molecule.

\section{Summary and conclusions}

In summary, we have shown rigorously how the mode-dependent scattering
cross sections and rates in a lattice can be determined using the
extended AGF method.~\citep{ZYOng:JAP18_Tutorial,ZYOng:PRB18_Atomistic}
This AGF-based approach allows us to treat scattering problems in
which additional degrees of freedom from an external scatterer are
present. Its utility is demonstrated with two examples: (1) the monoatomic
harmonic chain with a single isotopic impurity, and (2) the (10,10)
CNT with an encapsulated C60 molecule. In the first example, we derive
the exact analytical expressions for the scattering cross section
and rate. To first order in perturbation, the latter is identical
to the result in Ref.~\citep{PAllen:PRB22_Heat}. In the second example,
we illustrate how the scattering cross section varies with the properties
of the phonon mode (phonon branch and frequency), and show how phonon
modes with radial atomic motion are more strongly scattered by the
C60 molecule. 

This approach allows us to compute the phonon-defect scattering rates
which can be used as inputs in semiclassical phonon transport models.
It can also be used to study how the atomic structure of a defect
affects the scattering cross section of different phonon modes and
thus enable the more controlled use of nanostructuring and lattice
defects in phonon engineering for thermoelectric applications.~\citep{ZLiu:MRSB18_Nano}
With some straightforward modification, this approach can also be
extended to tightbinding lattice models to determine the scattering
cross sections associated with perturbations by external degrees of
freedom, because of the close analogy~\citep{JSWang:PRB06_Nonequilibrium,JSWang:EPJB08_Quantum,JSWang:FrontPhys14_NEGF}
between the AGF method and commonly used quantum transport models
for electrons.~\citep{DSFisher:PRB81_Relation,PAKhomyakov:PRB05_Conductance} 

\begin{acknowledgments}
The author acknowledges funding support from the Agency for Science,
Technology and Research (A{*}STAR) of Singapore with the Manufacturing,
Trade and Connectivity (MTC) Programmatic Grant ``Advanced Modeling
Models for Additive Manufacturing'' (Grant No. M22L2b0111) and from
the Polymer Matrix Composites Program (SERC Grant No. A19C9a004).
\end{acknowledgments}

\appendix

\section{The extended Atomistic Green's Function method~\label{sec:ExtendedAGFMethod}}

For the convenience of the reader and the sake of completeness, we
give a brief description of the extended AGF method, in which the
transmission and reflection matrices can be calculated, largely following
our earlier papers on extending the AGF method~\citep{ZYOng:JAP18_Tutorial,ZYOng:PRB18_Atomistic}.
Here, we discuss only the transmission matrices, which are relevant
to Eq.~(\ref{eq:TransmissionMatrix}). 

In the AGF setup in Fig.~\ref{fig:LayeredSystem}, the system is
partitioned into three subsystems: (1) the left lead, (2) the device
region, and (3) the right lead. Each lead consists of a semi-infinite
one-dimensional array of principal layers while the device region
corresponds to layers $n=1$ to $N-1$. In Fig.~\ref{fig:LayeredSystem},
the left and right leads are materially identical with the same interlayer
spacing $a_{\text{L}}=a_{\text{R}}$ and interatomic forces. 

We can write the mass-normalized force-constant matrix $\mathbf{H}$,
which represents the harmonic coupling of the entire system in Fig.~\ref{fig:LayeredSystem}
and has the block-tridiagonal structure, as 
\begin{equation}
\mathbf{H}=\left(\begin{array}{ccccccc}
\ddots & \ddots\\
\ddots & \boldsymbol{H}_{\text{L}}^{00} & \boldsymbol{H}_{\text{L}}^{01}\\
 & \boldsymbol{H}_{\text{L}}^{10} & \boldsymbol{H}_{\text{L}}^{00} & \boldsymbol{H}_{\text{LC}}\\
 &  & \boldsymbol{H}_{\text{CL}} & \boldsymbol{H}_{\text{C}} & \boldsymbol{H}_{\text{CR}}\\
 &  &  & \boldsymbol{H}_{\text{RC}} & \boldsymbol{H}_{\text{R}}^{00} & \boldsymbol{H}_{\text{R}}^{01}\\
 &  &  &  & \boldsymbol{H}_{\text{R}}^{10} & \boldsymbol{H}_{\text{R}}^{00} & \ddots\\
 &  &  &  &  & \ddots & \ddots
\end{array}\right)\ ,\label{eq:SystemForceConstantMatrix}
\end{equation}
where $\boldsymbol{H}_{\text{C}}$, and $\boldsymbol{H}_{\text{CL}}$
($\boldsymbol{H}_{\text{CR}}$) are respectively the mass-normalized
force-constant submatrices corresponding to the device region, and
the coupling between the device region and the semi-infinite left
(right) lead. We can associate each layer in Fig.~\ref{fig:LayeredSystem}
with a block row in $\mathbf{H}$. The block row submatrices $\boldsymbol{H}_{\alpha}^{00}$
and $\boldsymbol{H}_{\alpha}^{01}$, where $\alpha=\text{L}$ and
$\alpha=\text{R}$ for the left and right lead, respectively, describe
the lead phonons. If we set the layers to be large enough so that
only adjacent layers can couple, then $\boldsymbol{H}_{\alpha}^{00}$
denotes to the intralayer force-constant matrix for each layer while
$\boldsymbol{H}_{\alpha}^{01}$ ($\boldsymbol{H}_{\alpha}^{10}$)
denotes the interlayer harmonic coupling between each layer and the
layer to its right (left) in the lead. Here, $\alpha$ is used as
the dummy variable for distinguishing the leads, with $\alpha=\text{L}$
and $\alpha=\text{R}$ corresponding to the left and right leads,
respectively. We also maintain the conceptual distinction between
the left-lead and right-lead force-constant matrices of $\boldsymbol{H}_{\alpha}^{00}$
and $\boldsymbol{H}_{\alpha}^{01}$ even though $\boldsymbol{H}_{\text{L}}^{00}=\boldsymbol{H}_{\text{R}}^{00}$
and $\boldsymbol{H}_{\text{L}}^{01}=\boldsymbol{H}_{\text{R}}^{01}$
in our method for computing the scattering cross section.

We note here that in spite of the infinite number of layers making
up the system, only a finite set of unique force-constant matrices
($\boldsymbol{H}_{\text{C}}$, $\boldsymbol{H}_{\text{CL}}$, $\boldsymbol{H}_{\text{CR}}$,
$\boldsymbol{H}_{\text{L}}^{00}$, $\boldsymbol{H}_{\text{L}}^{01}$,
$\boldsymbol{H}_{\text{R}}^{00}$ and $\boldsymbol{H}_{\text{R}}^{01}$)
are needed as inputs for the AGF calculation because the leads are
made up of identical layers and the Hermiticity of $\mathbf{H}$ implies
that $\boldsymbol{H}_{\text{LC}}=(\boldsymbol{H}_{\text{CL}})^{\dagger}$
and $\boldsymbol{H}_{\text{RC}}=(\boldsymbol{H}_{\text{CR}})^{\dagger}$,
and $\boldsymbol{H}_{\alpha}^{01}=(\boldsymbol{H}_{\alpha}^{10})^{\dagger}$.
For the sake of convenience, we also represent the $N-1$ layers in
the device region by a single matrix $\boldsymbol{H}_{\text{C}}$. 

In principle, the system dynamics are determined by the infinitely
large $\mathbf{H}$ in Eq.~(\ref{eq:SystemForceConstantMatrix}).
However, for the \emph{effective} dynamics at a fixed frequency $\omega$,
the lattice dynamics problem becomes more tractable as we need only
to project the dynamics onto a finite portion of the system,~\citep{JSWang:EPJB08_Quantum,NMingo:Springer09}
corresponding to layers $n=0$ to $N$ in Fig.~\ref{fig:LayeredSystem},
to determine phonon scattering by the device region. Hence, we can
use the submatrices in Eq.~(\ref{eq:SystemForceConstantMatrix})
to construct the \emph{effective} harmonic matrix for this subsystem,~\citep{JSWang:EPJB08_Quantum}
\begin{equation}
\mathbf{H}^{\prime}=\left(\begin{array}{ccc}
\boldsymbol{H}_{\text{L}}^{\prime} & \boldsymbol{H}_{\text{LC}}^{\prime} & 0\\
\boldsymbol{H}_{\text{CL}}^{\prime} & \boldsymbol{H}_{\text{C}}^{\prime} & \boldsymbol{H}_{\text{CR}}^{\prime}\\
0 & \boldsymbol{H}_{\text{RC}}^{\prime} & \boldsymbol{H}_{\text{R}}^{\prime}
\end{array}\right)\ ,\label{eq:ProjectedForceConstantMatrix}
\end{equation}
where $\boldsymbol{H}_{\text{L}}^{\prime}=\boldsymbol{H}_{\text{L}}^{00}+\boldsymbol{H}_{\text{L}}^{10}\boldsymbol{g}_{\text{L},-}^{\text{ret}}\boldsymbol{H}_{\text{L}}^{01}$
and $\boldsymbol{H}_{\text{R}}^{\prime}=\boldsymbol{H}_{\text{R}}^{00}+\boldsymbol{H}_{\text{R}}^{01}\boldsymbol{g}_{\text{R},+}^{\text{ret}}\boldsymbol{H}_{\text{R}}^{10}$
correspond to the left and right edges, respectively while $\boldsymbol{H}_{\text{C}}^{\prime}=\boldsymbol{H}_{\text{C}}$
and $\boldsymbol{H}_{\text{CL/CR}}^{\prime}=\boldsymbol{H}_{\text{CL/CR}}=(\boldsymbol{H}_{\text{LC/RC}}^{\prime})^{\dagger}$.
The retarded surface Green's functions $\boldsymbol{g}_{\text{L},-}^{\text{ret}}(\omega)$
and $\boldsymbol{g}_{\text{R},+}^{\text{ret}}(\omega)$ are given
by \begin{subequations} 
\begin{equation}
\boldsymbol{g}_{\alpha,-}^{\text{ret}}=[(\omega^{2}+i0^{+})\boldsymbol{I}_{\alpha}-\boldsymbol{H}_{\alpha}^{00}-\boldsymbol{H}_{\alpha}^{10}\boldsymbol{g}_{\alpha,-}^{\text{ret}}\boldsymbol{H}_{\alpha}^{01}]^{-1}\ ,\label{eq:RetardedLeftSurfaceGF}
\end{equation}
\begin{equation}
\boldsymbol{g}_{\alpha,+}^{\text{ret}}=[(\omega^{2}+i0^{+})\boldsymbol{I}_{\alpha}-\boldsymbol{H}_{\alpha}^{00}-\boldsymbol{H}_{\alpha}^{01}\boldsymbol{g}_{\alpha,+}^{\text{ret}}\boldsymbol{H}_{\alpha}^{10}]^{-1}\ .\label{eq:RetardedRightSurfaceGF}
\end{equation}
\label{eq:AllRetardedSurfaceGF}\end{subequations} Physically, Eq.~(\ref{eq:RetardedLeftSurfaceGF})
denotes the retarded surface Green's function for a decoupled semi-infinite
lattice extending infinitely to the left (denoted by the ``-'' in
the subscript of $\boldsymbol{g}_{\alpha,-}^{\text{ret}}$) while
Eq.~(\ref{eq:RetardedRightSurfaceGF}) denotes the corresponding
surface Green's function for a decoupled semi-infinite lattice extending
infinitely to the right (denoted by the ``+'' in the subscript of
$\boldsymbol{g}_{\alpha,+}^{\text{ret}}$). In addition, the advanced
surface Green's functions can be obtained from the Hermitian conjugates
of Eq.~(\ref{eq:AllRetardedSurfaceGF}), i.e., $\boldsymbol{g}_{\alpha,-}^{\text{adv}}=(\boldsymbol{g}_{\alpha,-}^{\text{ret}}){}^{\dagger}$
and $\boldsymbol{g}_{\alpha,+}^{\text{adv}}=(\boldsymbol{g}_{\alpha,+}^{\text{ret}})^{\dagger}$. 

Given Eqs.~(\ref{eq:ProjectedForceConstantMatrix}) and (\ref{eq:AllRetardedSurfaceGF}),
we can assemble the pieces needed for the phonon scattering calculations.
The first piece is the corresponding Green's function for Eq.~(\ref{eq:ProjectedForceConstantMatrix})
$\boldsymbol{G}^{\text{ret}}=[(\omega^{2}+i0^{+})\mathbf{I}^{\prime}-\mathbf{H}^{\prime}]^{-1}$,
where $\mathbf{I}^{\prime}$ is an identity matrix of the same size
as $\mathbf{H}^{\prime}$; the $\boldsymbol{G}^{\text{ret}}$ matrix
can be partitioned into blocks in the same manner as $\mathbf{H}^{\prime}$,
i.e.,
\begin{equation}
\boldsymbol{G}^{\text{ret}}=\left(\begin{array}{ccc}
\boldsymbol{G}_{\text{L}}^{\text{ret}} & \boldsymbol{G}_{\text{LC}}^{\text{ret}} & \boldsymbol{G}_{\text{LR}}^{\text{ret}}\\
\boldsymbol{G}_{\text{CL}}^{\text{ret}} & \boldsymbol{G}_{\text{C}}^{\text{ret}} & \boldsymbol{G}_{\text{CR}}^{\text{ret}}\\
\boldsymbol{G}_{\text{RL}}^{\text{ret}} & \boldsymbol{G}_{\text{RC}}^{\text{ret}} & \boldsymbol{G}_{\text{R}}^{\text{ret}}
\end{array}\right)\ .\label{eq:FiniteGreensFunction}
\end{equation}
The next step is the computation of the advanced and retarded Bloch
matrices~\citep{TAndo:PRB91_Quantum,PAKhomyakov:PRB05_Conductance,ZYOng:PRB15_Efficient}
of the left and right leads, $\boldsymbol{F}_{\alpha}^{\text{adv/ret}}(+)$
and $\boldsymbol{F}_{\alpha}^{\text{adv/ret}}(-)$, which describe
the bulk translational symmetry of the layers along the direction
of the heat flux and can be computed directly from the formulas: \begin{subequations}
\begin{equation}
\boldsymbol{F}_{\alpha}^{\text{adv/ret}}(+)=\boldsymbol{g}_{\alpha,+}^{\text{adv/ret}}\boldsymbol{H}_{\alpha}^{10}\ ,\label{eq:RightGoingBlochMatrix}
\end{equation}
\begin{equation}
\boldsymbol{F}_{\alpha}^{\text{adv/ret}}(-)^{-1}=\boldsymbol{g}_{\alpha,-}^{\text{adv/ret}}\boldsymbol{H}_{\alpha}^{01}\ .\label{eq:LeftGoingBlochMatrix}
\end{equation}
\label{eq:BlochMatrices}\end{subequations} The bulk eigenmodes for
the lead, which describe the mode-dependent atomic displacement in
each layer at frequency $\omega$, can be determined directly from
the Bloch matrices, \begin{subequations} 
\begin{equation}
\boldsymbol{F}_{\alpha}^{\text{adv/ret}}(+)\boldsymbol{U}_{\alpha}^{\text{adv/ret}}(+)=\boldsymbol{U}_{\alpha}^{\text{adv/ret}}(+)\boldsymbol{\Lambda}_{\alpha}^{\text{adv/ret}}(+)\ ,\label{eq:RightGoingModes}
\end{equation}
\begin{equation}
\boldsymbol{F}_{\alpha}^{\text{adv/ret}}(-)^{-1}\boldsymbol{U}_{\alpha}^{\text{adv/ret}}(-)=\boldsymbol{U}_{\alpha}^{\text{adv/ret}}(-)\boldsymbol{\Lambda}_{\alpha}^{\text{adv/ret}}(-)^{-1}\ ,\label{eq:LeftGoingModes}
\end{equation}
\label{eq:BlochMatrixEigenmodes}\end{subequations} where $\boldsymbol{U}_{\alpha}^{\text{ret}}(+)$
{[}$\boldsymbol{U}_{\alpha}^{\text{ret}}(-)${]} is a matrix with
its column vectors corresponding to the rightward-going (leftward-going)
extended or rightward (leftward) decaying evanescent modes and has
the form $\boldsymbol{U}_{\alpha}^{\text{ret}}=(\boldsymbol{e}_{1}\boldsymbol{e}_{2}\ldots\boldsymbol{e}_{N})$
where $\boldsymbol{e}_{n}$ is a normalized eigenvector of the Bloch
matrix in the $n$-th column of $\boldsymbol{U}_{\alpha}^{\text{ret}}(+)$
{[}$\boldsymbol{U}_{\alpha}^{\text{ret}}(-)${]}. Similarly, $\boldsymbol{U}_{\alpha}^{\text{adv}}(-)$
{[}$\boldsymbol{U}_{\alpha}^{\text{adv}}(+)${]} is a matrix with
its column vectors corresponding to rightward-going (leftward-going)
extended or leftward (rightward) decaying evanescent modes. The matrix
$\boldsymbol{\Lambda}_{\alpha}^{\text{adv/ret}}(+)$ {[}$\boldsymbol{\Lambda}_{\alpha}^{\text{adv/ret}}(-)${]}
is a diagonal matrix with matrix elements of the form $e^{ik_{n}a}$
where $k_{n}$ is the phonon wave vector corresponding to the $n$-th
column eigenvector in $\boldsymbol{U}_{\alpha}^{\text{adv/ret}}(+)$
{[}$\boldsymbol{U}_{\alpha}^{\text{adv/ret}}(-)${]}. In a complex
pseudo-1D system such as the carbon nanotube, we can also attach to
each eigenmode an extra label corresponding to the branch index $\nu$
so that each eigenmode has a label $\nu_{n}k_{n}$.

The final piece of ingredient needed for the phonon scattering calculations
is the diagonal velocity matrix~\citep{PAKhomyakov:PRB05_Conductance,JSWang:EPJB08_Quantum}
\begin{align}
\boldsymbol{V}_{\alpha}^{\text{adv/ret}}(+)= & \frac{ia_{\alpha}}{2\omega}[\boldsymbol{U}_{\alpha}^{\text{adv/ret}}(+)]^{\dagger}\boldsymbol{H}_{\alpha}^{01}[\boldsymbol{g}_{\alpha,+}^{\text{adv/ret}}-\nonumber \\
 & (\boldsymbol{g}_{\alpha,+}^{\text{ret/adv}})^{\dagger}]\boldsymbol{H}_{\alpha}^{10}\boldsymbol{U}_{\alpha}^{\text{adv/ret}}(+)\ ,\label{eq:RightGoingVelocityMatrix}
\end{align}
which has group velocities of the eigenvectors in $\boldsymbol{U}_{\alpha}^{\text{adv/ret}}(+)$
as its diagonal elements. Likewise, $\boldsymbol{V}_{\alpha}^{\text{adv/ret}}(-)$
is defined as
\begin{align}
\boldsymbol{V}_{\alpha}^{\text{adv/ret}}(-)= & -\frac{ia_{\alpha}}{2\omega}[\boldsymbol{U}_{\alpha}^{\text{adv/ret}}(-)]^{\dagger}\boldsymbol{H}_{\alpha}^{10}[\boldsymbol{g}_{\alpha,-}^{\text{adv/ret}}-\nonumber \\
 & (\boldsymbol{g}_{\alpha,-}^{\text{ret/adv}})^{\dagger}]\boldsymbol{H}_{\alpha}^{01}\boldsymbol{U}_{\alpha}^{\text{adv/ret}}(-)\ .\label{eq:LeftGoingVelocityMatrix}
\end{align}
For evanescent modes, the group velocity is always zero while for
propagating modes that contribute to the heat flux, the group velocity
is positive (negative) in $\boldsymbol{V}_{\alpha}^{\text{ret}}(+)$
and $\boldsymbol{V}_{\alpha}^{\text{adv}}(-)$ {[}$\boldsymbol{V}_{\alpha}^{\text{ret}}(-)$
and $\boldsymbol{V}_{\alpha}^{\text{adv}}(+)${]}. In addition, we
define the diagonal matrices $\widetilde{\boldsymbol{V}}_{\alpha}^{\text{adv/ret}}(+)$
and $\widetilde{\boldsymbol{V}}_{\alpha}^{\text{adv/ret}}(-)$ in
which their nonzero diagonal matrix elements are the inverse of those
of $\boldsymbol{V}_{\alpha}^{\text{adv/ret}}(+)$ and $\boldsymbol{V}_{\alpha}^{\text{adv/ret}}(-),$
respectively. 

From Eqs.~(\ref{eq:AllRetardedSurfaceGF}) to (\ref{eq:LeftGoingVelocityMatrix}),
we construct the flux-normalized transmission matrix used in Eq.~(\ref{eq:TransmissionMatrix}):
\begin{align}
\boldsymbol{t}_{\text{RL}}= & \frac{2i\omega}{\sqrt{a_{\text{R}}a_{\text{L}}}}[\boldsymbol{V}_{\text{R}}^{\text{ret}}(+)]^{\nicefrac{1}{2}}[\boldsymbol{U}_{\text{R}}^{\text{ret}}(+)]^{-1}\nonumber \\
 & \times\boldsymbol{G}_{\text{RL}}^{\text{ret}}[\boldsymbol{U}_{\text{L}}^{\text{adv}}(-)^{\dagger}]^{-1}[\boldsymbol{V}_{\text{L}}^{\text{adv}}(-)]^{\nicefrac{1}{2}}\ .\label{eq:tmatrix_RL}
\end{align}
Each row of $\boldsymbol{t}_{\text{RL}}$ corresponds to either a
transmitted right-lead extended or evanescent mode. For an outgoing
evanescent mode, the row elements and group velocity, given by the
diagonal element of $\boldsymbol{V}_{\text{R}}^{\text{ret}}(+)$,
are zero. Conversely, each column of of $\boldsymbol{t}_{\text{RL}}$
corresponds to either an incoming left-lead extended or evanescent
mode, and the column elements and group velocity of the evanescent
modes, given by the diagonal element of $\boldsymbol{V}_{\text{L}}^{\text{adv}}(-)$,
are zero. If the $m$-th row and $n$-th column of $\boldsymbol{t}_{\text{RL}}$correspond
to extended transmitted and incoming modes, then $|[\boldsymbol{t}_{\text{RL}}]_{mn}|^{2}$
gives us the probability that the incoming left-lead phonon is transmitted
across the device region into the right-lead phonon. Similarly, we
can define the flux-normalized transmission matrix for phonon transmission
from the right to the left lead:

\begin{align}
\boldsymbol{t}_{\text{LR}}= & \frac{2i\omega}{\sqrt{a_{\text{L}}a_{\text{R}}}}[\boldsymbol{V}_{\text{L}}^{\text{ret}}(-)]^{\nicefrac{1}{2}}[\boldsymbol{U}_{\text{L}}^{\text{ret}}(-)]^{-1}\nonumber \\
 & \times\boldsymbol{G}_{\text{LR}}^{\text{ret}}[\boldsymbol{U}_{\text{R}}^{\text{adv}}(+)^{\dagger}]^{-1}[\boldsymbol{V}_{\text{R}}^{\text{adv}}(+)]^{\nicefrac{1}{2}}\ .\label{eq:tmatrix_LR}
\end{align}

\section{Generalization to two or three dimensions~\label{sec:2D3DGeneralization}}

Although the examples given in this paper are only applied to 1D (monotatomic
chain) or pseudo-1D systems (e.g. carbon nanotube), the generalization
of our method to 2D or 3D systems merits a more detailed discussion.
It should be noted that the original AGF method~\citep{WZhang:NHT07_Atomistic}
has been applied to the investigation of phonon transport in 2D and
3D systems~\citep{WZhang:JHT07_Simulation,DSingh:JHT11_Effect,SSadasivam:ARHT14_Atomistic}
and its extension, as described in Appendix~\ref{sec:ExtendedAGFMethod},
can be applied in a straightforward manner. The key difference between
a 1D system and its 2D or 3D counterpart is that in the latter, each
phonon eigenmode of the lead, from Eq.~(\ref{eq:BlochMatrixEigenmodes}),
is characterized by its transverse momentum ($k_{y},k_{z})$ in addition
to its longitudinal momentum $k_{x}$, frequency $\omega$ and branch
index $\nu$. Examples of the characterization of lead phonon eigenmodes
by their transverse momenta can be found in Refs.~\citep{ZYOng:PRB18_Atomistic,ZYOng:PRB20_Structure,QSong:PRB21_Evaluation,ZYOng:PRB24_Effect}. 

For simplicity, we will only sketch the 2D case as the 3D case can
be generalized from the 2D case. In a 2D system such as a graphene
sheet with a finite but sufficiently large geometrical width of $\mathcal{T}$
and coplanar with the $x-y$ plane, we impose periodic boundary conditions
in the transverse ($y$) direction and open boundary conditions in
the longitudinal ($x$) direction. We should more accurately characterize
the system as being \emph{quasi}-2D because of its finite width. When
$\mathcal{T}$ is sufficiently large, the computed scattering of cross
section is expected to approach the exact value.

In the case of a single-layer graphene, the leads are pristine graphene
sheets semi-infinite in the longitudinal direction while the device
is a graphene sheet of finite length with either a defect or has some
external scatterer (e.g. a functional group) attached to it. Hence,
in such a system, every extended phonon mode of the lead is a 2D phonon
mode and is characterized by a discrete transverse momentum $k_{y}$,~\citep{SSadasivam:ARHT14_Atomistic}
which is given by $2\pi n/\mathcal{T}$ for $n=0,\pm1,\pm2,\ldots$,
in addition to its longitudinal momentum $k_{x}$, which is a function
of $\omega$ and $k_{y}$. In a phonon branch with the linear dispersion
$\omega=ck$ where $c$ is the group velocity, the longitudinal momenta
$k_{x}$ of the eigenmodes at frequency $\omega$ would be discretized
and given by $k_{x}=\pm\sqrt{(\omega/c)^{2}-(2\pi n/\mathcal{T})^{2}}$
for $n=0,\pm1,\ldots,\pm N$ and $N=\lfloor\frac{\omega\mathcal{T}}{2\pi c}\rfloor$.
In the $\mathcal{T}\rightarrow\infty$ limit, we recover the idealized
2D system and the transverse and longitudinal momenta are continuous. 

After labeling each lead phonon eigenmode by its transverse momentum,
we can compute its scattering cross section using Eq.~(\ref{eq:CrossSectionFormula})
with the geometrical cross section $A$ replaced by $\mathcal{T}$.
In the case of a 2D system, we can rewrite Eq.~(\ref{eq:RationalizedTransmissionMatrix})
as:
\begin{equation}
\overline{\boldsymbol{t}}_{\text{RL}}(\omega)=\left(\begin{array}{ccc}
S(\nu_{1}\bm{k}_{1},\nu_{1}\bm{k}_{1}) & \cdots & S(\nu_{1}\bm{k}_{1},\nu_{\bar{l}}\bm{k}_{\bar{l}})\\
\vdots & \ddots & \vdots\\
S(\nu_{\bar{l}}\bm{k}_{\bar{l}},\nu_{1}\bm{k}_{1}) & \cdots & S(\nu_{\bar{l}}\bm{k}_{\bar{l}},\nu_{\bar{l}}\bm{k}_{\bar{l}})
\end{array}\right)\ ,\label{eq:RationalizedTransmission}
\end{equation}
where the momentum of the phonon mode is represented by a vector $\bm{k}=(k_{x},k_{y})$
instead of a scalar $k$, with $\bm{k}=\bm{k}_{n}$ and $1\leq n\leq\bar{l}$.
From Eq.~(\ref{eq:RationalizedTransmission}), we can determine the
scattering cross sections $\sigma(\nu_{1}\bm{k}_{1})$ to $\sigma(\nu_{\bar{l}}\bm{k}_{\bar{l}})$. 

\bibliography{PaperReferences}

\end{document}